\documentstyle[12pt,aaspp4] {article}

\def\etal{et al.\rm}
\def\f555w{{\it F555W}\/}

\righthead{Q0957 Precision Photometry}

\begin{document}

\title{Hourly Variability in Q0957+561}

\author{Wesley N. Colley and Rudolph E. Schild}
\affil{Harvard-Smithsonian Center for Astrophysics, 60 Garden Street, Cambridge
MA 02138}

\begin{abstract}

We have continued our effort to re-reduce archival Q0957+561 brightness
monitoring data and present results for 1629 $R$-band images using the methods
for galaxy subtraction and seeing correction reported previously (Paper I).
The new dataset comes from 4 observing runs, several nights apiece, with
sampling of typically 5 minutes, which allows the first measurement of the
structure function for variations in the $R$-band from timescales of hours to
years.  Comparison of our reductions to previous reductions of the same data,
and to $r$-band photometry produced at Apache Point Observatory (Colley \etal\
1999) shows good overall agreement.  Two of the data runs, separated by 417
days, permit a sharpened value for the time delay of 417.4 days, valid only if
the time delay is close to the now-fashionable 417-day value; our data do not
constrain a delay if it is more than three days from this 417-day estimate.
Our present results show no unambiguous signature of the daily microlensing,
though a suggestive feature is found in the data.  Both time delay measurement
and microlensing searches suffer from from the lack of sampling at half-day
offsets, inevitable at a single observatory, hence the need for round-the-clock
monitoring with participation by multiple observatories.

\end{abstract}


\section{Introduction}

Previous studies of the Q0957+561 A,B brightness fluctuations have indicated
the existence of daily brightness fluctuations.  If these are intrinsic to the
quasar, then the time delay can be determined to a precision of a fraction of a
day; however, if such fluctuations are at least in part due to microlensing,
they would probably signal the nature of a baryonic dark matter component.
 
A recent report by Colley and Schild (ApJ 518, 153, 1999) demonstrated
precision aperture photometry close to the photon limit.  The procedure
subtracts out the lens galaxy's light according to the galaxy's profile
determined recently from an HST image, and removes cross talk light from the A
and B image apertures.  We now apply this reduction to 1629 frames collected in
4 runs, with continuous monitoring for 10 hours/night over, typically, 5
nights.
 
Because previous analysis of these data showed that small brightness
fluctuations were observed during these runs, but because the effects of seeing
were known to limit conclusions about such low-amplitude fluctuations, we hoped
that re-reduction would clarify conclusions about time delay and microlensing.
In the first several sections to follow, new results unravel the systematic
effects that seeing has on the aperture photometry.  Next, the measured
brightness fluctuations give a structure function that describes the amplitude
of variations observed in this lensed/microlensed system on various timescales.
Final sections present time delay calculations and possible microlensing
results from two observing runs separated by 417 days.

\section{The Photometry}

For two decades, RS has monitored Q0957 and amassed a ``master dataset'' of the
variations of images A and B (Schild \& Thompson 1995 and earlier references
therein).  The data presented here come from all-night monitoring in one of two
similar observational configurations.  About half of the data, as in Paper I,
derives from unbinned 1k$\times$1k CCD images with pixel scales of one-third of
an arcsecond, exposure times of 450 seconds, read noise of $8e^-$, and FWHM
seeing of 1.5--2 arcseconds.  Some of the data come from otherwise similar
circumstances, but derive from a 2k$\times$2k CCD binned down to 1k$\times$1k;
here shorter exposure times are used, and the pixel scale is twice as great.
Both of these setups typically allow for 5 unsaturated comparison stars in the
$R$-band on each data-frame.  This paper presents the photometry from more than
1600 such frames.

To reduce 1600 frames, one requires a highly automated photometry reduction
code (one free of mouse clicks and manual entry of parameters).  In Paper I, we
detailed such a method, which is not only automated, but includes two basic
improvements upon the basic aperture photometry scheme historically employed to
reduce Q0957+561 data images: 1) use of HST imaging to subtract the galaxy, 2)
correction for ``cross talk'' between the narrowly separated ($6.1\arcsec$) A
and B images.  Typically the galaxy, whose core is inside the B aperture,
contributes 17\% to the B aperture and 2\% to the A aperture.  The cross talk
contributes typically 2\% to each image.  One could calibrate those out if the
seeing were steady, but poor seeing obviously introduces more cross talk
contamination, and has the curious effect of spreading much more galaxy light
into the A image.  These variable effects confuse the search for both intrinsic
QSO variation and microlensing.  Worse still, while the seeing introduces a
correlation in the A and B photometry (spreading more cross talk light into
both images from each other), it also introduces an {\it anti-}correlation due
to the galaxy, whose light is spread out from the B aperture and into the A
aperture.  To disentangle these effects, a complete galaxy subtraction and
measurement of cross talk is necessary.

Slight modifications to the photometry scheme in Paper I have been implemented,
principally for sake of speed and increased automation.  As before the
astrometry is accomplished via PSF fitting, which requires a model PSF.  The
model is empirical, built by stacking several standards.  The first pass
utilizes methods described by Alard \& Lupton (1998), but only even terms in
the polynomial expansion are used.  An explicitly even model PSF for each
standard can be fit to the standard itself to determine the astrometry of that
standard.  Shifting and stacking (by median) all of the standards forms the
model PSF.  This method has the advantages of being quite fast and of
eliminating cosmic rays automatically from the model PSF.  Other modifications
include increased sensitivity to erratic photometry in the standards (by simply
censoring 5-$\sigma$ outliers), and more careful measurement of crosstalk
(again by filtering out gross outliers).  Each of these methods increase the
stability of the photometry appreciably.

\section{The Effects of the Lens Galaxy}

\begin{figure}[t]
\plotone{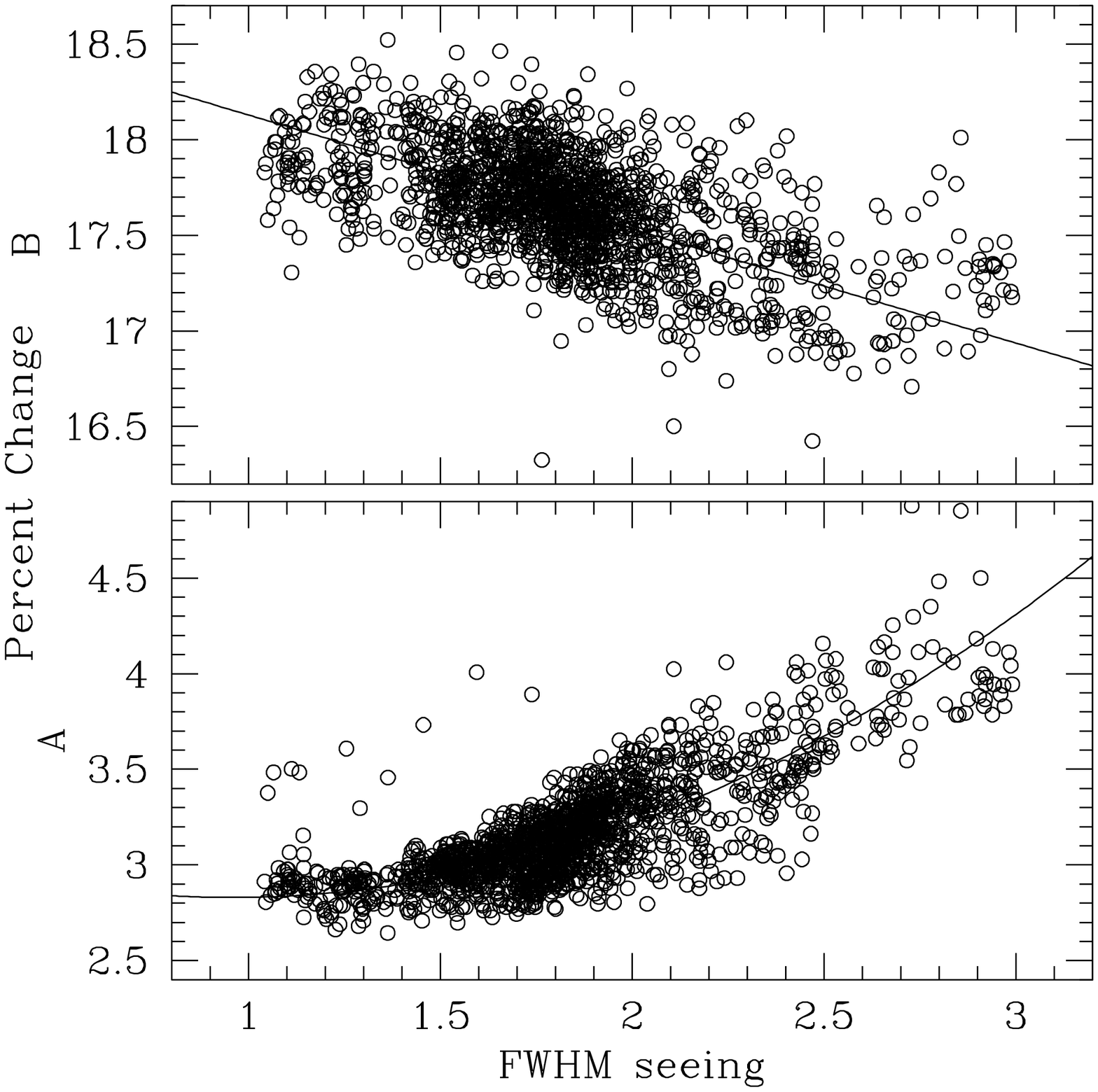}
\caption{Galaxy Contamination as a function of seeing.  Plotted is the
percentage of light in the A and B apertures due to the lens galaxy.
Contamination by the galaxy not only alters the average photometry of the QSO
images, but adds a component which is variable as a function of seeing.  This
variable component has an amplitude of up to 3\% from good ($1\arcsec$) to bad
($2.5\arcsec$) seeing, large enough to corrupt searches for few percent
brightness variations. As seeing deteriorates, light of the lens galaxy is
systematically scattered out of the image B aperture, which contains the
galaxy's nucleus, but scattered into the A aperture. The fitted curves are
discussed in section 3.}
\label{gal}
\end{figure}

Fig.~\ref{gal} shows the contribution of the lens galaxy's light to the A and B
measurement apertures as a function of seeing, as measured from all the image
frames of this data set. Plotted is the percentage of the signal in the
apertures as a function of FWHM seeing, measured from the stellar images on
the data frames. Because seeing effects originate in different levels of the
terrestrial atmosphere and do not have a single unique profile, adoption of a
simple FWHM parameter is only a first-order statistic and cannot fully describe
the variety of seeing profiles supplied by nature.

In the top panel of Fig.~\ref{gal} the typical percentage contribution to the
light of image B is 17.7 percent for average seeing.  The average brightness of
the B image during this time was 16.51 magnitudes.  The observed contribution
is equivalent to an 18.39 magnitude source in the measurement aperture. This is
in excellent agreement with the R = 18.34 magnitude correction to the lens
galaxy magnitude that has been historically adopted throughout the 20-year
Q0957 brightness monitoring project (Schild and Weekes, 1984, ApJ 277,
481). The historical correction was made from a compilation of aperture
photometry at optical and infrared wavelengths, and the known colors of an
elliptical galaxy at moderate redshift.  Published data tables of Schild and
colleagues usually had a correction for an R = 18.34 magnitude galaxy applied,
but no correction for the galaxy contribution to the A aperture and no aperture
cross talk corrections.

The bottom panel in Fig.~\ref{gal} shows the brightness contribution from the
lens galaxy to the aperture of the A image. For average seeing, this correction
is 3\%. Because this contribution originates in the previously unknown outer
profile of the lens galaxy, no correction for it has been made in the
historical brightness records (where bad seeing images were removed). This has
little effect on the historical record, since it just causes a slight reduction
in the measured brightness fluctuations of image A, and a slight 3\% correction
to the inferred A/B continuum brightness ratio.

Best fit parabolic curves have been overplotted on the Fig.~\ref{gal} data; we
give those fits below.  With seeing expressed as a FWHM (represented below as
$x$) in units of arcseconds, the fitted curves express the percent contribution
of the lens galaxy to the R magnitude in a $6\arcsec$ diameter A aperture
$\delta R(A)$ and B aperture $\delta R(B)$ as
\begin{equation}
\begin{array}{ll}
\delta R(A) = [3.154 - 0.6775x + 0.3542x^2]\%, \\
\delta R(B) = [18.725-0.5962x]\%.
\end{array}
\end{equation}
We found that the best fit quadratic term for B was very nearly zero, so we
omit that here.  These empirical low order fits apply only in the range shown
on the plots, and should not be used for seeing with FWHM less than one
arcsecond.  In such cases, the minimum value shown for the A contamination is
probably a quite good approximation, since the curve is obviously flattening by
one arcsecond anyway.  It seems, however, that the B contamination is still
growing as the seeing improves toward one arcsecond, and perhaps one could
extrapolate somewhat, but the B contamination would obviously level below some
FWHM; hence, correction for the galaxy in the B aperture during very good
seeing remains somewhat unresolved.

\section{The Aperture Corrections}

\begin{figure}[t]
\plotone{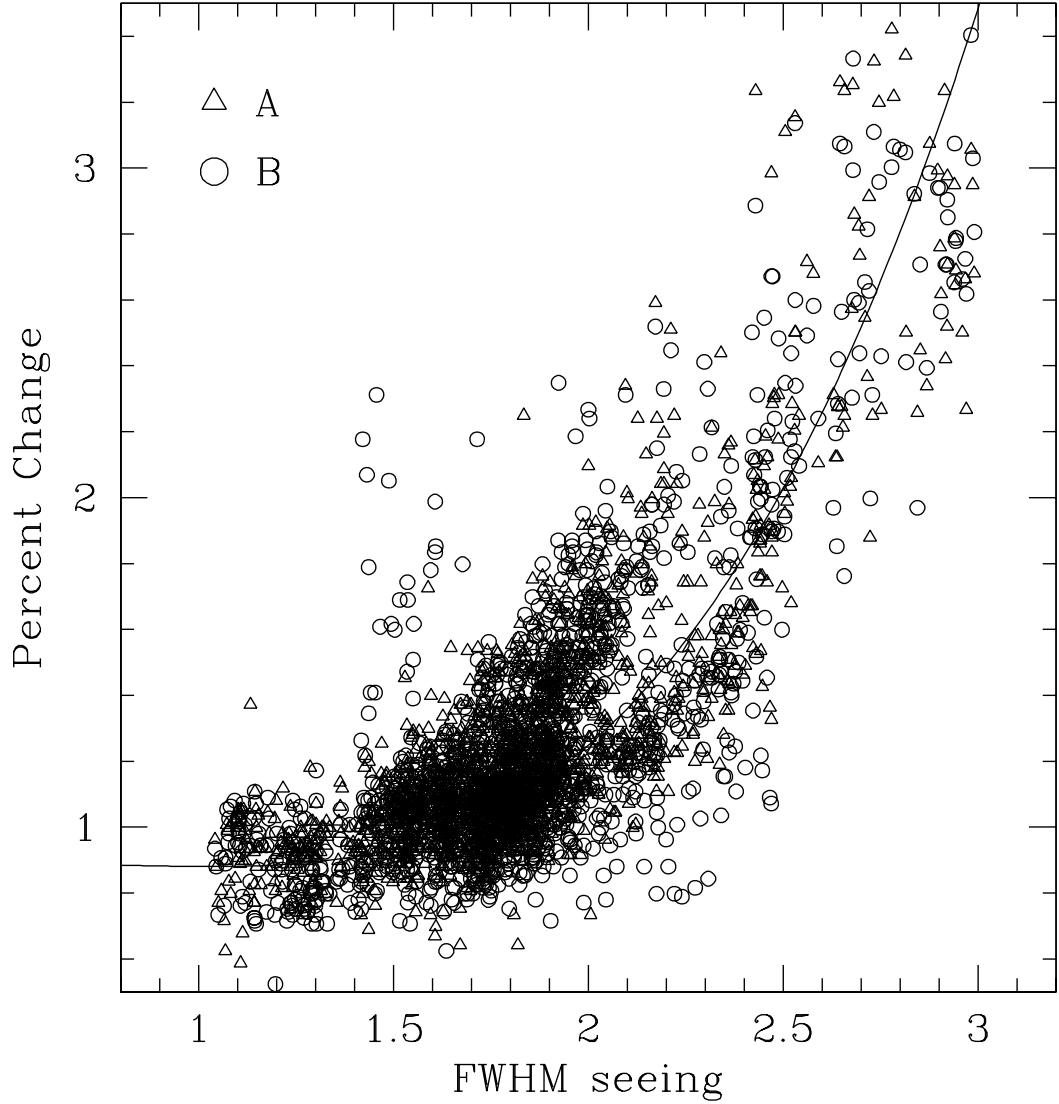}
\caption{Aperture crosstalk as a function of seeing. The crosstalk correction
for the A and B images on each data frame is determined from synthesized
apertures adjacent to 4 standard stars on each data frame, and averaged.  The
corrections are the same for the A and B apertures indicating that the star
images are symmetrical. The fit curve shown is discussed in Section 4.}
\label{seecro}
\end{figure}

As the seeing gets worse, relatively more light will spill over from quasar
image A into the adjacent B aperture (and vice versa) and cause systematic the
cross talk effects noted by Schild and Cholfin, 1986. This effect can be
corrected for by measuring the amplitude of the effect in nearby field
stars. Insofar as the telescope/camera optics produce no asymmetrical image
structure, the effect should be the same for the A and B apertures, as we find
here.  Thus Fig.~\ref{seecro} shows the aperture correction curve applicable to
both apertures, with the amplitude of the crosstalk expressed in percent as a
function of seeing FWHM.

The shape and nature of this curve will necessarily depend somewhat on the
optics of the telescope/camera and on the properties of the terrestrial
atmosphere above Mt Hopkins.  We find that a cubic curve represents the
crosstalk corrections quite well, and has the following form for $\delta R$ as
a function of the seeing FWHM ($x$):
\begin{equation}
\delta R = [0.7087 + 0.6320x - 0.7387x^2 + 0.2788x^3]\%
\end{equation}
Again, these curves should not be applied for seeing FWHM below one arcsecond,
for which the correction evidently flattens out at a value very close to
one percent.

Results in this and the previous section show two very different effects of
seeing in the two measurement apertures. The crosstalk correction is virtually
the same for the two apertures, but the galaxy correction opposes the crosstalk
in image B, while adding with the crosstalk for image A. Thus the B image data
are substantially more seeing independent than image A data.

Since most of the historical Q0957 data (from the ``master dataset'' of Schild
and Thomson [1995]) were taken with seeing between 1 and 2 arcseconds, the
corrections are always less than about a percent.  For seeing worse than 2.5
arcseconds, the observations have been heavily censored in the past, which
Fig.~\ref{seecro} demonstrates to be important since seeing effects can produce
photometric errors of several percent.

\section{Results: The Brightness Record}

\begin{figure}[p]
\plotone{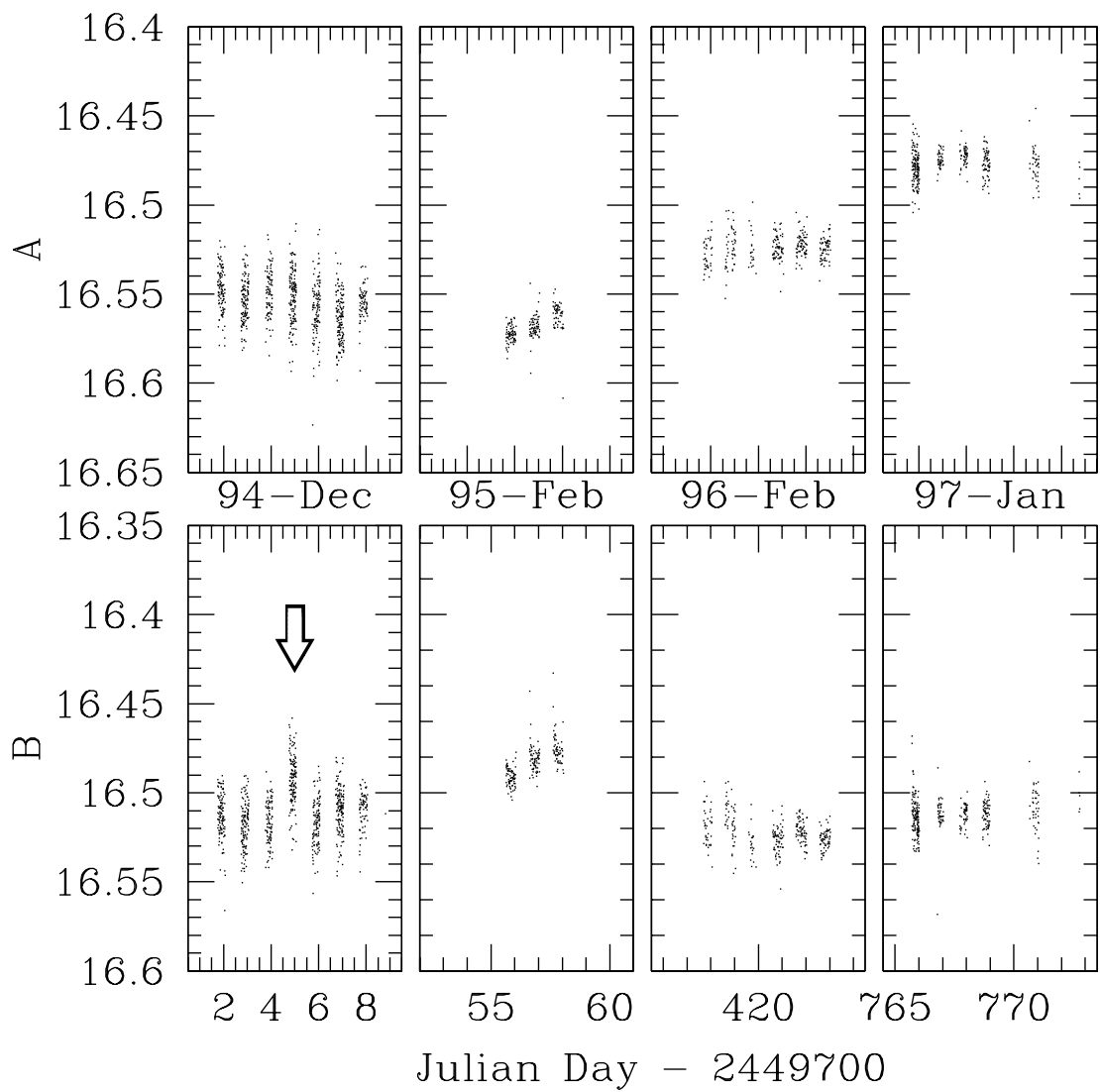}
\caption{Brightness record after corrections for the lens galaxy and aperture
crosstalk.  Each data point is for a single image frame, and the ``swarms''
give an idea of the spread of photometry image to image.  Despite the photon
noise scatter, variations night-to-night are visible.  One particularly
interesting feature is in the B data from 1994-December, pointed out by the
arrow, where for a single night the B image jumps by 2\%. Nightly averages for
this data are shown in Fig.~\ref{nightraw}b) and the hourly binned data for
94-Dec and 96-Feb are shown as Fig.~\ref{centfig})}
\label{alldata}
\end{figure}

The final brightness data after corrections for the lens galaxy and aperture
cross talk appear in Fig.~\ref{alldata}.  Plotted points are the result from
each data frame, so the amplitude of the ``swarm'' for each night relates to
the accuracy of the photometry. The accuracy is comparable to our results in
Paper I, where it was found that the photometry is within 15\% of the limit
imposed by Poisson statistics. Note that the swarm for the first observing run
in Dec. 1994 covering Julian Dates 2449702--08 has a larger amplitude than for
later runs; this is because the pixel size was larger by a factor of two on the
first observing run, so exposure times are four times shorter and the Poisson
noise greater. But although the noise per image frame was larger, there are
more of them and the overall precision of the photometry is about the same as
for subsequent observing runs.

Correlations in the A and B photometry could arise due to coherent
observational errors, such as crosstalk (Schild and Cholfin 1986), though no
significant correlation appear (verified by the Pearson's correlation
coefficient applied to raw, galaxy subtracted, and crosstalk corrected
photometry).  Although it might at first appear that correlated fluctuations
were seen in the Feb 1995 run, with both images brightening with time, note
that during this time the A image was unusually faint and the B image was
unusually bright.  So we conclude that by chance, both images were seen
brightening at the same time.

These brightness records give unambiguous evidence of daily brightness
fluctuations.  The best is the record for image B on night 2449704, which we
have highlighted with an arrow in Fig.~\ref{alldata} (the same feature is just
as obvious in the uncorrected photometry).  Here image B increased in
brightness by more than 2\% from the night before, and returned to its normal
level the night after.  We conclude that the quasar lens system is capable of
producing 2\% brightness fluctuations on a time scale of a day. In section 7
below we will compute the structure function for Q0957 brightness fluctuations
on long and short time scales.

\section{Comparison of Old and New Reductions}
 
The new reductions allow a comparison with the original CCD data record
published by Schild and colleagues over the years. Their data are available in
tabular form at URL:
http://cfa-www.harvard.edu/$\!\sim\!\,$rschild.
The data have also been published in Schild \& Thomson (1995, and references
therein).  Here we are concerned with the question of how well the new
reductions agree with the original published and tabulated data.  Recall that
the basic procedure in both the old and new reductions is aperture photometry,
and that the new reduction differs primarily in the treatment of subtraction of
the lens galaxy G1 and in the correction for aperture crosstalk.

\begin{figure}[t]
\plotone{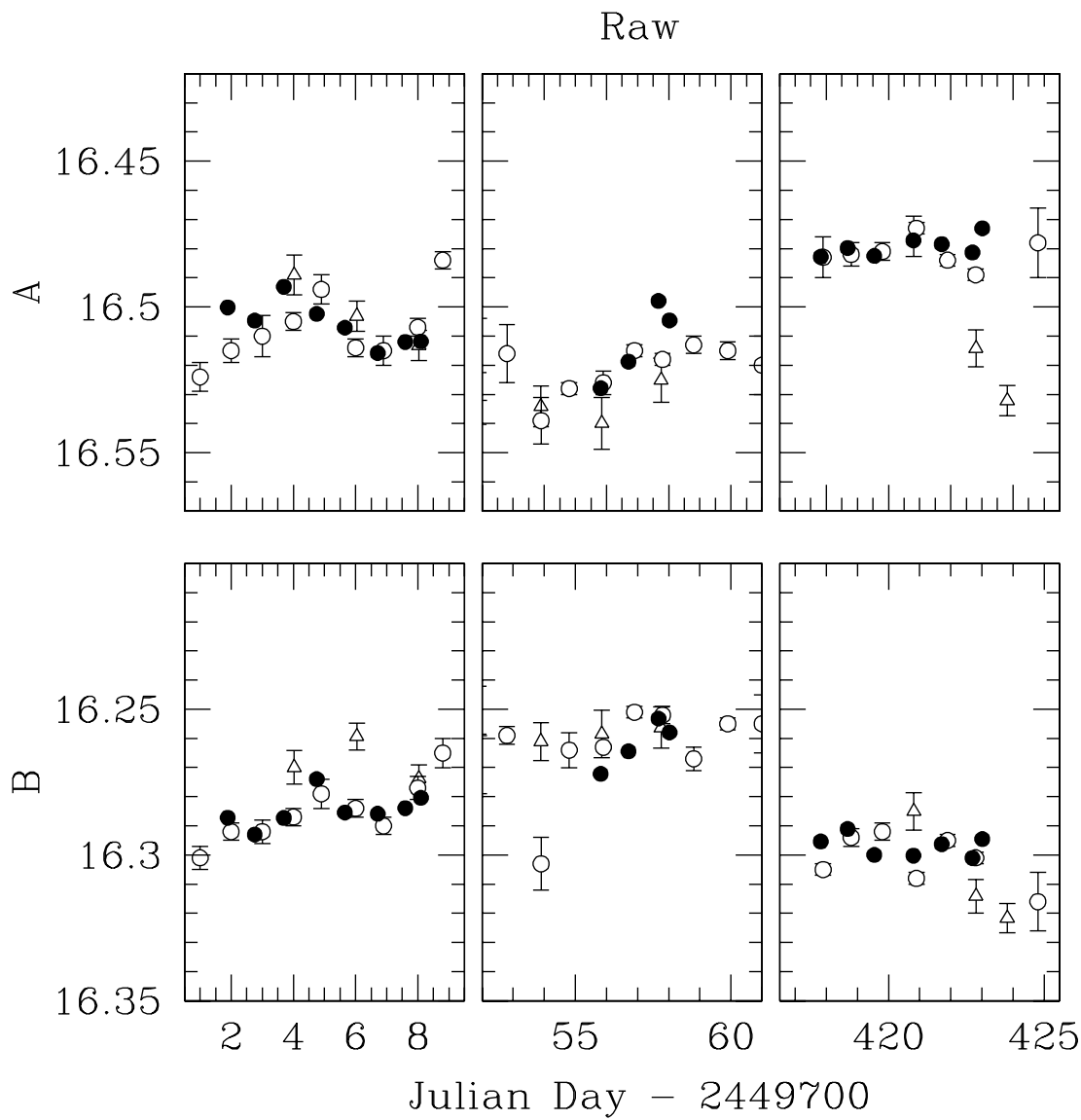}
\caption{a) Comparison of old and new reductions. Plotted as filled circles are
the new nightly brightness averages with no correction for galaxy subtraction
or aperture cross talk, to be comparable with the previously published
photometry.  The errorbars are often smaller than the plot symbols.  In open
circles, are points from Schild \& Thomson (1995); triangles are $r$-band data
from Princeton.  A single offset of 0.61 magnitudes has been applied to the
Princeton photometry, and a single offset of $-0.18$ magnitudes has been
applied to the Schild image B photometry, to compensate for his galaxy
subtraction. It may be seen that the old Schild reductions using IRAF software
produced results very similar to the new reductions.}
\label{nightraw}
\end{figure}

Plotted in Figs.~\ref{nightraw}a) and b) are the photometry from this work in
filled circles.  In open circles are plotted the previous reductions of these
data from Schild's master dataset.  In open triangles is the $r$-band
photometry from Apache Point (Colley \etal~1999).  Fig.~\ref{nightraw}a) shows
the comparisons with the raw aperture photometry in this work, while
Fig.~\ref{nightraw}b) gives the comparisons with the corrected photometry.

In Fig.~\ref{nightraw}a), no correction or shift has been applied to either the
old or new photometry for image A.  The overall agreement is obviously quite
good (almost always within 2\%).  Meanwhile, for image B, we have simply added
back in the galaxy light subtracted in Schild's master dataset (his 18.34
magnitude galaxy correction), and there is quite good agreement (rarely
exceeding more than 1\% error).  The reason the observation times appear
slightly different for these two datasets is that in the Schild dataset,
filtering for seeing removed large fractions of data from each run and hence
shifted the mean time of observation for each night.  If one linearly
interpolates the Schild data and compares directly to the photometry of this
work, one finds that the image A photometry averages just 3 mmag brighter than
Schild's, with a standard deviation of 8 mmag (statistically consistent with no
offset at all).  If one does the same for image B, one finds an average
discrepancy of 184 mmag with a standard deviation of 6 mmag.  This 184 mmag
means there is an additional 18.31 magnitude source in our raw B aperture, very
close to the 18.34 mag correction Schild made for the lens galaxy.

The [A,B] rms differences between the photometries of only [8,6] mmag show that
the nightly averages published by Schild and colleagues are confirmed, since
the published errors average about the same amount, which adds credence to many
microlensing conclusions reported based upon this data.  Note particularly from
Fig.~\ref{nightraw}a) that the discrepancies between the old and new reductions
are on a time scale of individual nights, which should not effect the
microlensing trends on larger timescales proposed from wavelet analysis by
Schild (1999).

Fig.~\ref{nightcro}b) gives the same data from Schild, but plots our corrected
photometry.  In previous sections, we discussed the contributions of the galaxy
and cross talk during typical seeing.  For aperture A the lens galaxy has
contributes approximately 3\% during typical seeing, while the cross talk
between apertures contributes another 1.2\%.  Our corrected photometry is
accordingly 44 mmag fainter than the Schild photometry with an rms of 7 mmag (a
slight improvement over the comparison with raw photometry).  The image B
photometry shows a 40 mmag deficit to the Schild photometry with an rms of 5
mmag (also a slight improvement over the raw photometry).  The improvement in
rms is perhaps expected, because of the filtering for good seeing done by
Schild.

Schild and Thomson (1995) report an internal photometric error of 10 and 12
mmag for A and B respectively during this epoch.  Therefore, the agreement of
the Schild photometry and current photometry at the 5 mmag level, is about as
good as one could expect.  It is perhaps noteworthy that the agreement exists
despite completely different software for the aperture photometry (IRAF dophot
vs. IDL programming from scratch) and completely different methods for the
relative photometry (pencil and calculator vs. Honeycutt [1992] matrix
reduction).

Because we expect to continue re-reducing archival data frames with our
improved data reductions, it is worthwhile summarizing the differences
between the new and old reductions.  Relative to the old reductions, the new
results for image [A,B] are [44, 40] mmag fainter than the tabulated Schild
photometry, primarily because of the new corrections for aperture crosstalk
but also partly due to improved knowledge of the effects of the lens galaxy.
Thus to correct our old photometry to the new system, one would add [.044,
.040] mag to the table entries in the Schild master data set. The true
errors of the photometry for the dates we have compared are [7,5] mmag.

For reference, the open triangles in Figs.~\ref{nightraw}a) and b) give the APO
$r$-band photometry where applicable (Colley \etal\ 1999), with arbitrary
offsets.  While the gross features of the photometry are reproduced, the
agreement is not as good, but this might be expected because 1) $r$ is a
different filter, 2) no correction or censoring for seeing effects, and 3) the
APO photometry derives from only a handful of frames on a given night (hence
the larger errorbars), compared with several tens of frames per night at
Mt. Hopkins.

\section{The Structure Function for Q0957 Brightness Fluctuations}

For many purposes related to the discussion of the Q0957 time delay and
interpolation of the data set for missing observations, it is useful to know
the structure function for the brightness fluctuations. The structure function,
which is the expected variance as a function of temporal separation, can be
approximated as a power law $V^2 = 10^a\cdot\tau^b$.  For our corrected
photometry, we find $a = -4.5; b = 0.54$, in rough agreement with values found
by Colley \etal\ (1999) for $r$-band photometry of $a = -4.1; b = 0.47$, and by
PRH for $R$-band photometry of $a_A \approx -0.34; b_A \approx 0.54$ and $a_B
\approx -0.22; b_B \approx 0.68$ for the A and B images respectively.  The
difference in the A and B power laws in PRH shows that measurement of the
structure function is perhaps not the most stable process (even perhaps not the
best way to describe QSO variation), and suggests that we should not be
troubled by the smallish discrepancies between each of these estimates.
Fortunately, within reason, the details of the structure function have little
bearing on the PRH statistic itself.

Fig.~\ref{strfn} contains a plot of the structure function for the fluctuations
averaged for images A and B.  Note that Schild (1996) has found from wavelet
analysis that the A and B images have equal brightness fluctuations on time
scales from days to 2 months, and Pelt et al (1998) have shown that on time
scales of decades, the fluctuations are larger in B, presumably because of its
larger optical depth to microlensing.

Fig.~\ref{strfn} shows the structure function for time scales of less than a
day for the first time.  Note that there is no obvious departure from the power
law from the smallest to the longest scales.  For time scales of a day, the
results are consistent with the frequent remark by Schild (1996) that this
quasar's daily brightness fluctuations are typically of order a percent (0.6
percent according to this plot).

\section{Time Delay Estimates}

Because two of the observing runs, December 1994 and February 1996 were
separated by 417 days, the presently favored value of the time delay
(Colley \etal\ 1999, Pelt \etal\ 1998), it should be possible to determine
the gravitational lens time delay to a fraction of a night.  We have
completed two time delay calculations for the new data.  For both of the
calculations, to be discussed below, the data have been binned by one hour.

Our first calculation, shown as a dotted line in Fig.~\ref{chi2}, is based upon
simple linear interpolation of the data and is equivalent to the calculation by
Kundi\'c et al (1997, ApJ 482, 75), in which errors and values are linearly
interpolated between the nearest points, and a simple $\chi^2$ statistic is
computed.  This autocorrelation calculation produces a network of minima
separated by 1.0 days, with the deepest minimum occurring for 417.3 days.

A time delay calculation using the Press, Rybicki, and Hewitt (1992)
calculation is shown as the solid line in Fig.~\ref{chi2}.  This calculation
uses the Fig.~\ref{strfn} structure function to estimate the permitted range of
brightness values in place of linear interpolation.  It also shows a network of
minima separated by 1.0 days with the deepest minimum at 417.5 days.

These minima at $n + 1/2$ days can be understood easily in terms of the
statistics used to compute the best-fit time delay, both of which prefer not to
overlap data.  The linear method prefers no overlap for a simple reason.
Because the data are binned by one hour, the actual number of data in the first
and last bin varies according to chance.  When a small number of data occurs in
those bins, the errorbars are, of course, larger by root-n, than those of
typical bins.  Hence the linearly interpolated errors are larger by the same
factor and are thus far more tolerant in the no-overlap regions.  The unusually
large errorbars at the end points also explains why the absolute $\chi^2$
minimum for the linear method is less than one.

The PRH method prefers no overlap for a different reason, but with similar
results.  The PRH method, of course, finds that the structure of variations of
A and B independently match the structure function (Fig.~\ref{strfn})
measured from the A and B variations.  It tries to keep A and B independent
rather than fighting to find the correct overlap.  This propensity to find the
gaps was recognized as a possible weakness of PRH by its authors in the
original paper.

While both of these methods have proven excellent for determining the time
delay for well-sampled data on different timescales (Colley \etal\ 1999), both
show weakness for intermittently sampled data, inevitable at a single
observatory at mid latitudes. Because of this complication we conclude that our
program of intensive monitoring over many nights separated by the time delay
does not allow us to sharpen the time delay significantly, and we do not claim
that our overall best fit value 417.4 days is an improvement over other recent
determinations.  This problem motivates the need for round-the-clock monitoring
of the QSO.

\section{Agreement of the Time Shifted Brightness Curves}

Fig.~\ref{centfig} contains the December 1994 A brightness record (filled
circles) and the shifted B brightness record (open circles).  At bottom the
image B photometry has been shifted by the best-fit time delay from the linear
interpolation $\chi^2$ method; at top the PRH fit is shown.  We show both,
because of the qualitatively different appearance of the fits, and because we
have no predisposition to favor either method particularly.  For reasons
discussed in the previous section, we might be inclined to place more faith in
the PRH method for producing the best overall fit, because it is less subject
to the edge effects due to binning.

In both panels of Fig.~\ref{centfig}, there does seem to be a qualitative
agreement between the behavior in A and B.  Particularly, there seems to be a
coherent wavy behavior in both A and B.  Some encouragement that the signal is
real lies in the fact that night to night neither A nor B shows systematic
behavior which is obviously spurious.  Namely, neither A nor B is always
increasing, decreasing or inflecting in the same way on each night.  Such
behavior would lead to suspicion that the waviness was an observational
artifact.  Without ``filling in the gaps,'' however, the actual validity of
these fits is moot, and we are left with only one certainty, that the QSO often
varies by about a percent within a given night.  This variation means that with
round-the-clock observations, highly precise measurement of the time-delay
would be possible, and any daily microlensing residing in the signal should
become apparent.

Looking at Fig.~\ref{centfig}, it is hard to find any definitive microlensing
signal, but most suggestive is the apparent gap at JD 2449705.8.  While the
rest of the curve (in the top panel) seems to obey a coherent pattern, this
last interface looks slightly off.  This is hardly irrefutable evidence for
microlensing, but again, with round-the-clock monitoring, we would be able to
be much more definitive if such an event arose.  Furthermore, a null result
would be of some interest: quasar microlensing searches for the missing mass
have been producing exclusion diagrams for possible MACHO masses (Schmidt and
Wambsganss 1998, Refsdal et al 1999).  The intermittency of microlensing and
clumping of caustics (Wambsganss, Witt, and Schneider 1992) are complications,
however, and a detection such as the Dec 1994--March 1996 event (Schild and
Gibson 1999) is challenging, not so much to observe but to confirm, because of
the low amplitudes of the events.

\section{Summary and Conclusions}

Using improved photometric methods as described in Paper I, we have begun our
program of re-reduction of archival Q0957 CCD images acquired over the last 20
years.  We report herein results for 4 data runs when the quasar was
continuously observed with a 1.2m telescope for typically 6 consecutive nights,
including observing runs in Dec 1994 and Feb 1996 which are separated by 417,
the currently favored time delay between images A and B (Kundi\'c \etal\ 1997).

For the new data we show how the contribution of the lens galaxy varies with
the seeing.  During good seeing the lens galaxy contribution to the A aperture
is nearly constant at 2.5\% and appears to maximize at about 18.5\% in the B
aperture.  As seeing deteriorates beyond $\mbox{FWHM} = 2.5\arcsec$, the A
aperture contribution increases by 1\% while the B aperture contribution
decreases by 1\%.

For the same data we evaluate the contribution of A-B aperture cross talk and
find that significant deterioration occurs in both apertures for seeing with
FWHM greater than $1.8\arcsec$.  The deviation from average due to this error
source reaches 2\% during $\mbox{FWHM} = 3\arcsec$ seeing. Thus we find that
for deteriorating seeing, galaxy contamination adds to aperture crosstalk in
aperture A and compensates for half the cross talk in aperture B.  These
effects can very significantly affect aperture photometry for the system, but
historically they have been only qualitatively understood.

For the fully corrected data we present hourly binned brightness curves that
show significantly detected variations on time scales of hours.  The quasar can
produce 2\% brightness fluctuations on the time scales of 24 hours, and 1\%
brightness fluctuations within individual nights of continuous observation.

Comparison of our new photometry with the previous reductions in the master
data set (Schild and Thomson 1995,
http://cfa-www.harvard.edu/$\!\sim\!\,$rschild)
yields excellent agreement.  The new raw photometry agrees with the historical
data within an rms of about 7 mmag.  The new corrected photometry agrees within
slightly better rms limits, after an offset of about +42 mmag is applied to the
historical photometry.

With two data runs separated by 417 days, the currently favored time delay, one
can search for a more precise delay, and for any daily microlensing residual.
Two different time delay estimation procedures favor values near half-day
values.  These estimators, as would most, tend to find gaps where there is no
data overlap between the two runs.  The best microlensing candidate in these
runs is a significant discrepancy on day 2449705.8 (Fig.~\ref{centfig}), which
is unfortunately straddling a half-day gap.  The tendancy for time delay
estimators to find the half-day gaps where either A or B could not be observed
due to daylight, and the lack of conclusiveness available for microlensing due
to such gaps necessitates round-the-clock monitoring of Q0957+561.

\newpage

\addtocounter{figure}{-1}

\begin{figure}[t]
\plotone{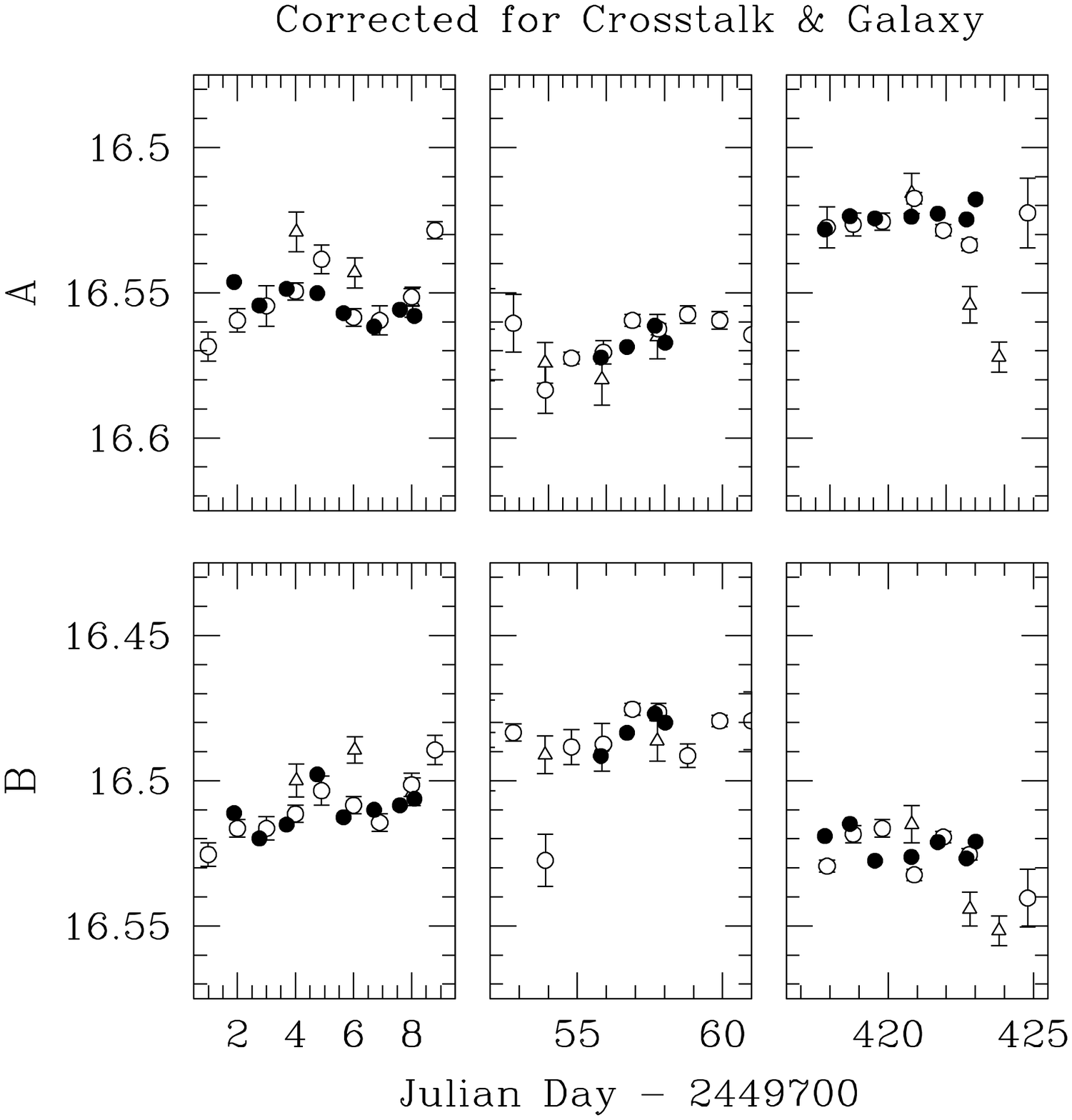}
\caption{b) As with Fig. \ref{nightraw}a), except the filled circles
represent photometry which has been corrected for the galaxy and for
crosstalk.  An arbitrary overall offset has been allowed for A and B, since the
photometry from Schild \& Thomson, and from Apache Point has not had neither
the galaxy subtracted nor crosstalk corrected.}
\label{nightcro}
\end{figure}

\begin{figure}[t]
\plotone{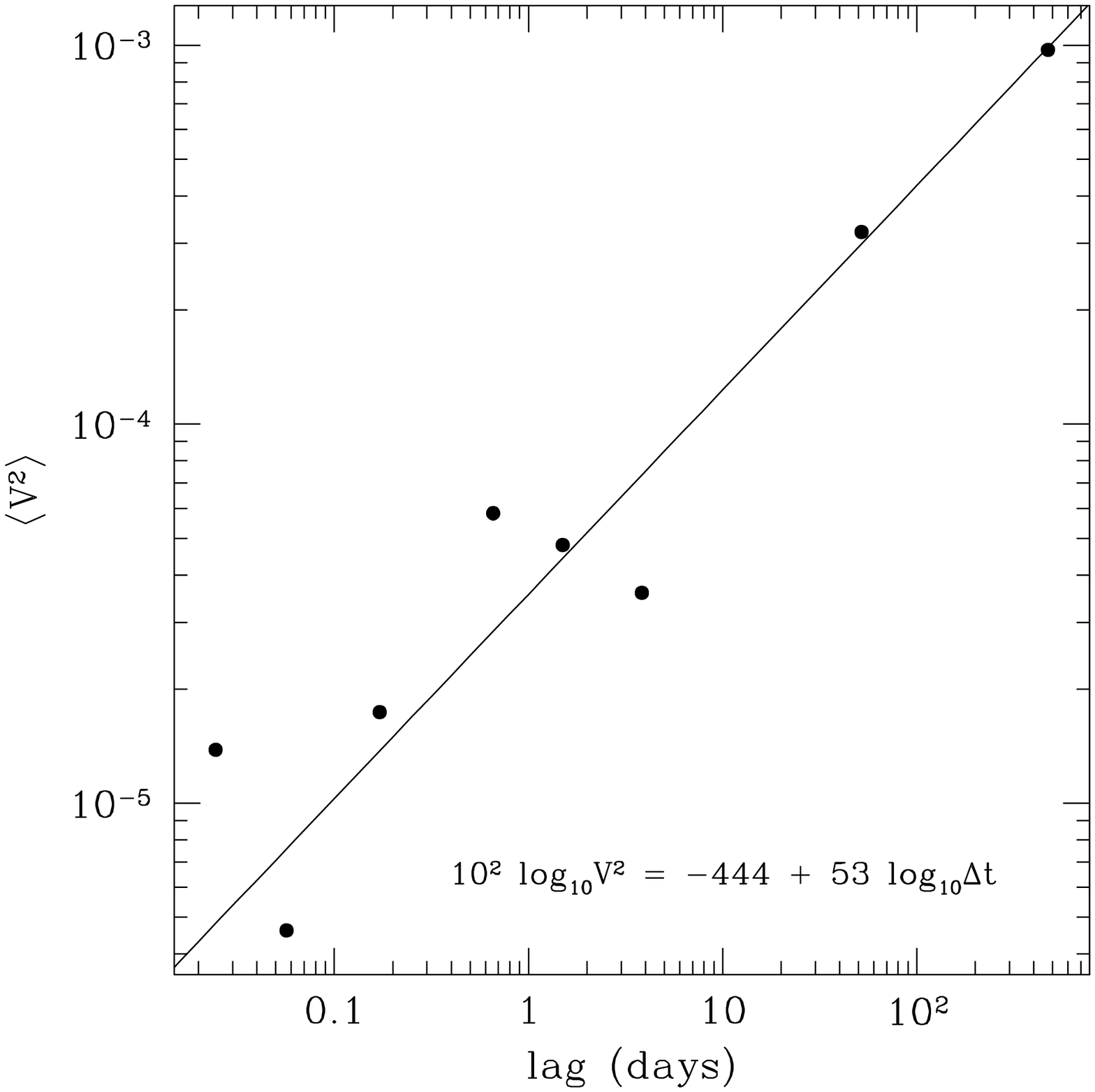}
\caption{The structure function for brightness fluctuations observed in the
lensed quasar system, as defined in Section 7. This structure function
shows that brightness fluctuations are of order a percent for 1-day
lags. Of course both microlensing and intrinsic quasar brightness
fluctuations contribute to the fluctuation amplitude.}
\label{strfn}
\end{figure}

\begin{figure}[t]
\plotone{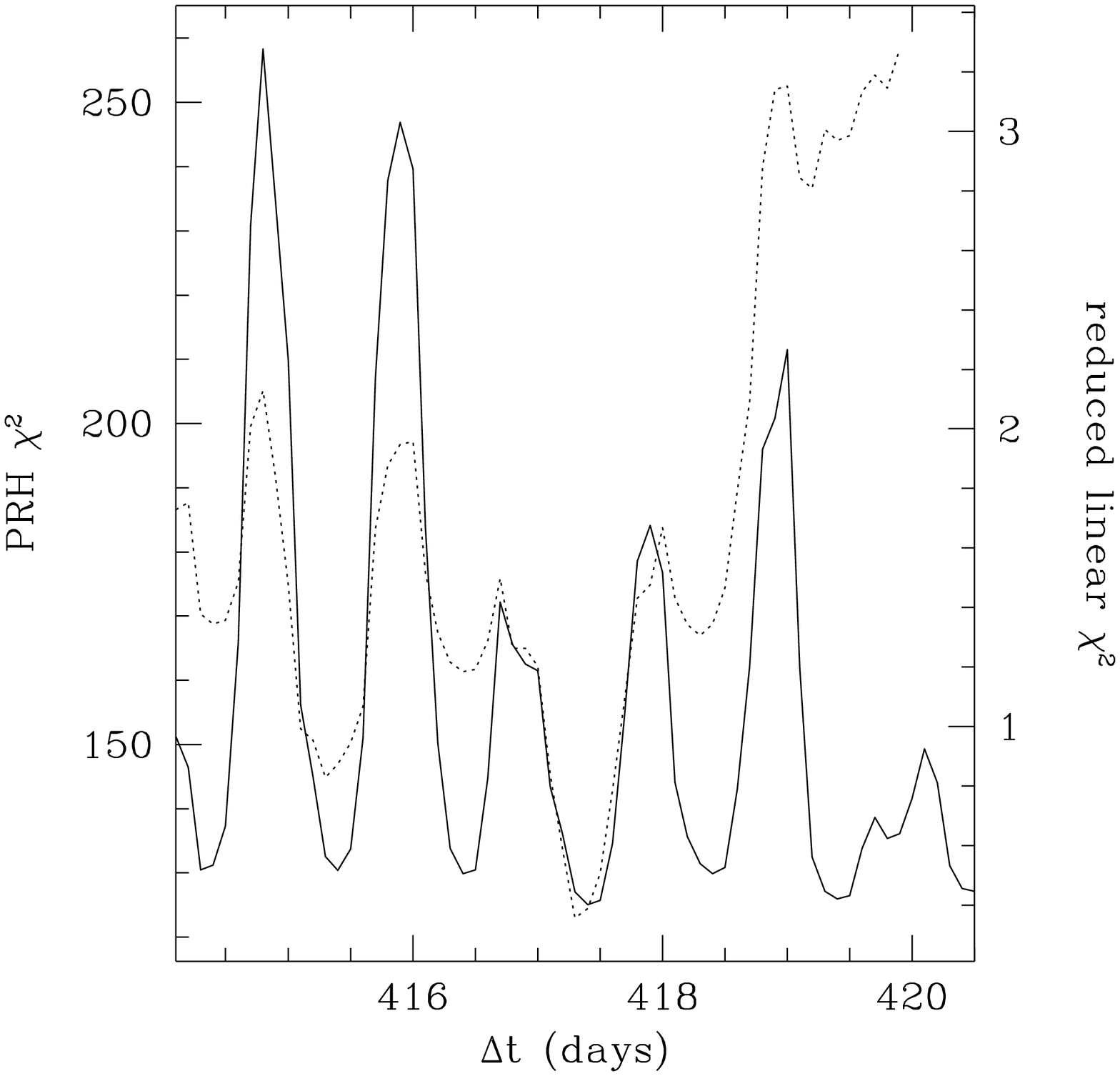}
\caption{Measurement of Time Delay from the Dec. 1994 image A data and the
Feb. 1996 image B data, binned by one hour (see Fig.~\ref{centfig}).  Plotted
are the $\chi^2$ figures of merit from two methods, the Press Rybicki \& Hewitt
(1992) method (solid curve) and a linear interpolation method (dotted curve)
introduced by Kundi\'c \etal\ (1997).  The methods show minima at 417.5 and
417.3 days, respectively.  However, both methods also show a troubling sequence
of minima at offsets of $\sim (n + 1/2)\mbox{~days}; n \in J$.  These minima
show up where there is no overlap in the data, due to the sun.  Obviously full
24-hour coverage would remedy this situation.}
\label{chi2}
\end{figure}

\begin{figure}[p]
\plotone{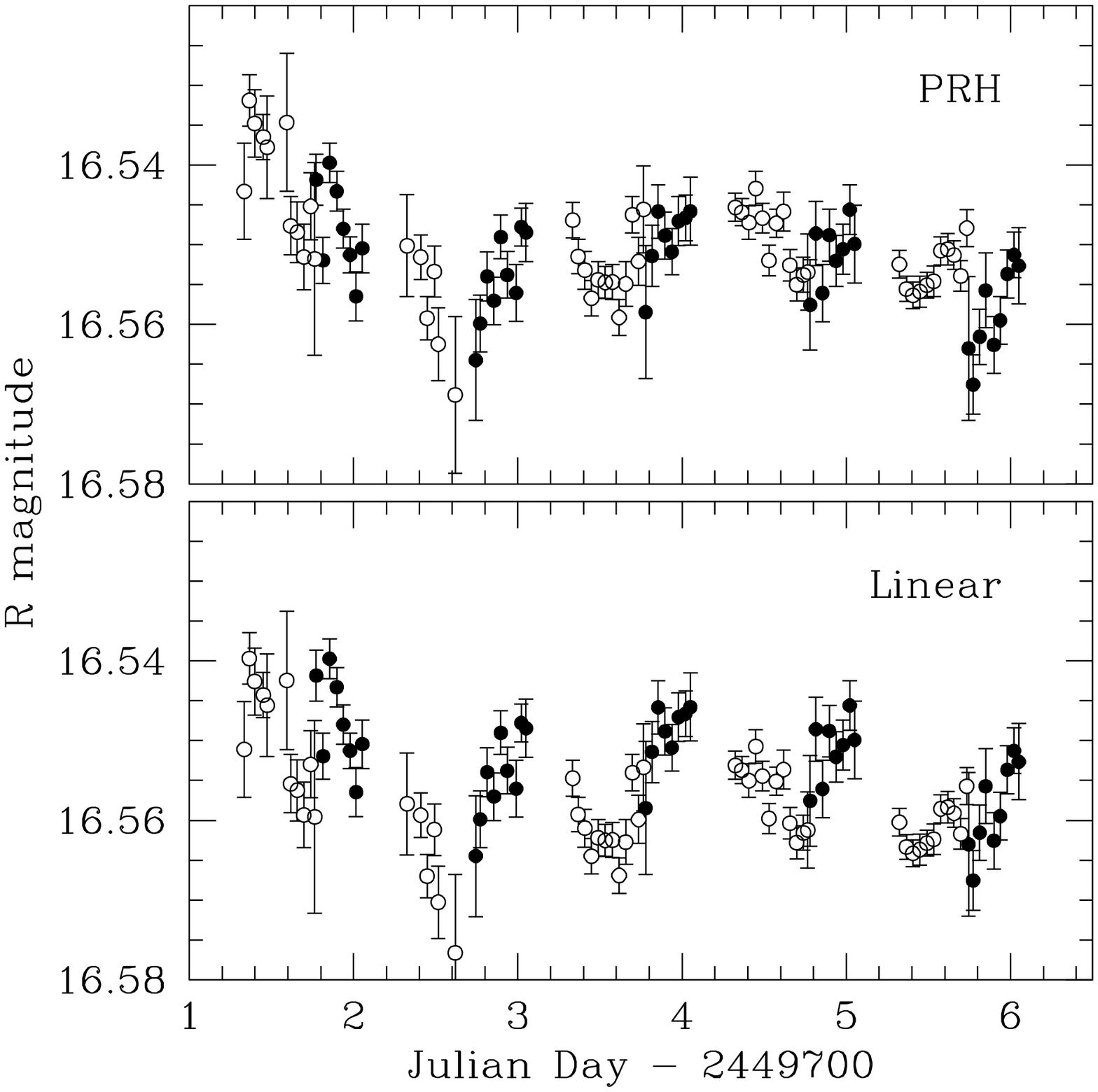}
\caption{Image A and shifted Image B Data (binned by one hour) in filled and
open points respectively.  At top, the time delay and offset have been
determined by the PRH method, at bottom by a simple linear interpolation scheme
(see Fig.~\ref{chi2}).  Brightness fluctuations of order 1\% can be seen during
some nights.  Such fluctuations should allow determination of the time-delay
with a precision of one hour.  After such a determination, subtraction of the A
light curve from the very precisely shifted B light curve should reveal any
short time-scale microlensing.  Microlensing on a time scale of a day might be
implied by the discontinuity at 2449705 days.  Such an event would imply very
small MACHO's ($\sim 10^{-5}M_\odot$) in the lens galaxy halo.}
\label{centfig}
\end{figure}

\end{document}